\documentclass[twocolumn]{revtex4}
\usepackage{graphicx,epsfig}
\usepackage{amsmath}
\usepackage{amsfonts}
\usepackage[colorlinks=true,breaklinks=True]{hyperref}
\usepackage{subfig}

\usepackage{fancyhdr}
   
\begin{document}


\pagestyle{fancy}
\lhead{\bf Phase space analysis of the young stellar component of the Radcliffe Wave}
\rhead{ }


\title{Phase space analysis of the young stellar component of the Radcliffe Wave}
\author{J. Donada and F. Figueras}
\affiliation{Institut de Ci\`encies del Cosmos (ICCUB), Universitat de Barcelona (IEEC-UB), Mart\'i i Franqu\`es 1, E-08028 Barcelona, Spain.}
\email{donada@ieec.cat}

\begin{abstract}

{\bf Abstract:} The Radcliffe Wave is a galactic-scale structure recently proposed by J. Alves et al. (\cite{alves}: Nature, 2019). The authors propose that various molecular complexes in the solar environment follow a specific alignment and displacement that make them worthy of a common origin and evolution. In this work, we first collected and analyzed the population of very young stars and open clusters around this structure. The criteria for cross-matching these star-forming tracers with the identified Radcliffe Wave cloud complexes have been defined and applied, all based on the quality of the available astrometric and photometric data. We performed a first characterization of the structure and kinematic properties of the young stellar population linked to this wave. Our conclusions, although very preliminary, are: 1) we have identified 13 open clusters, each of them physically linked to a Cloud Complex, which are probable members of the Radcliffe Wave;  2)   The OB field stars do not present the elongated structure that departs from an straight line at the Sun position observed in the Cloud Complexes; 3) the vertical motion of 11 CC-OCs members
associated with the Wave is not contradictory with the behaviour
expected from a simple model of harmonic motion in the
vertical direction, and 4) the orbits back on time neither
suggest an origin associated to a point nor to a straight
line in the $XZ$ plane.

\end{abstract}

\maketitle


\section{Introduction}
\label{intro}

For more than 150 years astronomers have reported and studied the structure of the Gould Belt. This is a system in the solar neighbourhood, of size about 500pc from the Sun, formed by young stars, interstellar dark clouds, molecular clouds, dust, and neutral HI. Excellent reviews on the characterization of this structure can be found in Poppel \cite{poppel} and Palou\v{s} and Ehlerová \cite{palous}.   To date, there has been no consensus on the possible model that justifies the presence of this structure, and even its existence has been questioned. Thanks to the great quality astrometric data provided by {\it Gaia} in the last two years, the community has embarked on a thorough review of this structure in the 6-D phase space. From that, we expect more reliable models to account for its origin and evolution. By using the {\it Gaia} second data release (DR2), Dzib et al. \cite{dzib} review the mean distances and motions to the star-forming regions defining the Gould Belt. Their results, although consistent with a mean expansion velocity of the system of about  $2.5 \pm 0.1 km \cdot s^{-1}$, also reflect the existence of some young stars with large  peculiar velocities. More recently, in 2019, by revisiting the 3-D structure of
all local cloud complexes, Alves et al. (\cite{alves}) have proposed a change of paradigm. They propose the existence of a narrow and coherent large structure of dense gas with a wave shape of size about 2.7 kpc, aligned  from the first to the third galactic quadrant, named the Radcliff Wave (hereafter RW). This structure is proposed to include Orion A stellar complex, one of the most significant star forming regions defining the Gould Belt (hereafter GB). Next to this elongated structure of the RW, the existence of a second linear structure near the solar neighbourhood has been proposed. It is named as “split”, it is of about 1kpc size, and is proposed to contain the Sco-Cen star forming region (see Lallement et al. 2019, \cite{lallement}). Therefore, even with new {\it Gaia} data, the controversy surrounding the existence of the Gould Belt remains. Maybe the GB could be a simple projection effect of two linear cloud complexes as seen from the Sun,  one of them being Orion in the Radcliffe Wave (RW) and the other Sco-Cen complex in the "split" structure. 

In this work we aim to provide new insights into this scenario. On the one hand, we will study the structure in the configuration space defined by two tracers, the very young OB field stars and the extremely young open clusters (hereafter OCs). Thanks to the {\it Gaia} data (DR2 and eDR3 releases) we know the position of these objects with unprecedented precision up to about 2-3 kpc from the Sun. This position will be compared with the location of the cloud complexes (hereafter CCs) recently published by Alves and collaborators (\cite{alves}). As a second objective, we intend to define, for the first time, its vertical motion and their orbital trace-back on time trajectories of the complexes associated with the Radcliffe Wave. This will be done by associating the CCs with the young open star clusters, a link that will allow us to assign a 3-D motion to the CCs and, from that, a trace back orbital analysis using a realistic Galactic potential. We want to answer the question of whether the waveform of the RW in the direction perpendicular to the galactic plane is, or is not, maintained over time.

In \cite{alves} this RW structure has been detected from the gas and using only the CCs. Our present work aims to analyze whether this structure can be also well identified using young stellar tracers. To this end, in Section 2 we present the available data of two of the well identified stellar tracers, that is: the OB field stars recently collected by Pantaleoni-González et al. 2021 (\cite{pantaleoni}) and the OB stars from Xu et al. catalogue (\cite{xu}), and also the new catalogue of young open clusters recently published by Castro-Ginard et al. (\cite{castro}). It is in this section where we evaluate the over/under-densities of the stellar component associated with the Radchiffe Wave and the "split" structure and the Gould Belt, among others.  In Section 3 we define, justify and apply the set of criteria that allow us to  link these stellar tracers to the gas. Once associated, and as a third goal, we have used the publicly available python code for orbit computation to trace back on time the location of these structures in the last 30 million years. Finally, in Section 4 we discus our preliminary results and the work we intend to develop in the near future.

\section{The gas and stellar tracers}
\label{Section1}

As mentioned, up to now,  the Radcliffe Wave has only been studied using cloud complexes, but neither through open clusters nor through the  OB field stars. In order to provide new insights on the RW, first its members (as many as possible) must be identified. Therefore, in this study the young stellar component of the RW is characterised by means of three tracers: cloud complexes (Alves et al. catalogue \cite{alves}), field OB stars (catalogues by Xu et al. \cite{xu}, and by Pantaleoni-González et al. \cite{pantaleoni}) and open clusters (by Castro-Ginard et al. \cite{castro}). For each tracer sample, the available data, its quality and the spatial distribution is analyzed. As mentioned, the kinematic data is needed to trace the dynamical evolution of this structure and CCs do not have this information. On the contrary, the open cluster (OC) catalogue contains mean proper motions and radial velocities for the majority of objects. For OB field stars, proper motions are available from {\it Gaia}-eDR3 and some radial velocity data could be collected through SIMBAD Database.

\subsection{The Cloud Complexes}
Alves et al. provide a catalogue (\cite{alves}) of the distances to 326 of the major star-forming gas clouds in the local interstellar medium (CCs). Although this catalogue is available and the fitting algorithm to identify the RW members is described in the paper, we do not have detailed information on the membership assignment.  The distribution of these 326 CCs is shown as the red point series in the right column of Fig. \ref{fig:mapesxy}.

\subsection{The young  open clusters}

The open cluster catalogue used is the one provided by Castro-Ginard et al. (\cite{castro}), containing 1867 open clusters (OCs). According to Torra et al. (\cite{torra}), the estimated age of the Gould Belt system is in the interval 30-60 Myr, and the RW is assumed to be defined by star forming regions of a very young age (\cite{alves}). Therefore, as a first task only OCs younger than 30 Myr have been selected. They are 349 in total, accounting for approximately 20\% of the catalogue objects with the youngest one having an assigned age of 1.6 Myr. The distribution on the Galactic plane XY map of this young sample is shown as gray dots in the top left graphic of Fig. \ref{fig:mapesxy}. For 199 OCs out of the sample of 349 young OCs, the catalogue provides {\it Gaia}-eDR3 distances, proper motions and radial velocity data. This data is very precise as it has  been obtained as the mean of  the data available for its members. Once an OC will be associated to its corresponding CCs in Sect. \ref{link}, this kinematic information will allow us to trace back the orbit of the CCs assigned to the RW. 

\subsection{The OB stars in the field }

\begin{figure*}[hbt!]
\centering
\includegraphics[width=0.83\textwidth]{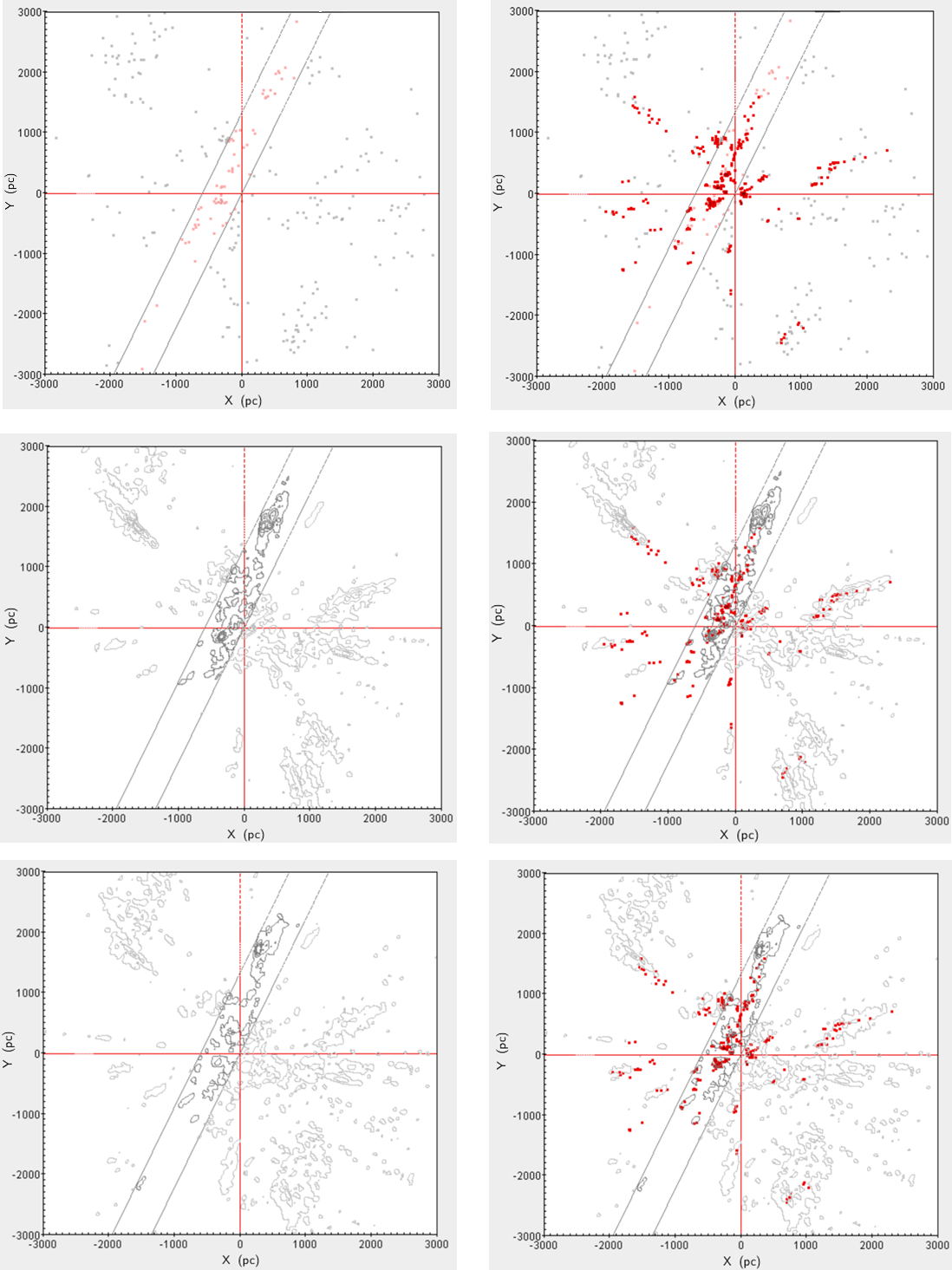}
\caption{XY map of the Radcliffe Wave tracers (left column) and their same map with the cloud complexes of Alves et al. catalogue (\cite{alves}) overplotted as red points (right column). The top panel corresponds to the open clusters in Castro-Ginard et al. catalogue (\cite{castro}) younger than 30Myr,  and the mid and bottom panels are the density maps of the OB stellar population: the mid panel corresponds to the Pantaleoni-Gonz\'alez et al. OB stars catalogue  (\cite{pantaleoni}, using the distances estimated with J. Maíz Apellániz's Bayesian formalism: \cite{maiz2001} and \cite{maiz2005}), and the bottom panel corresponds to the Xu et al. OB stars catalogue  (\cite{xu}). The straight inclined grey lines delineate a region containing the Radcliffe Wave. Both lines are parallel and tilted 30º relative to the Y axis in the first galactic quadrant, one passing through the Sun’s position (origin) and the other being horizontally translated 600pc to the left.}
\label{fig:mapesxy}
\end{figure*}

The studied OB stars tracers belong to two recently published catalogues. One is the Pantaleoni-González et al. catalogue (\cite{pantaleoni}), hereafter named as ALS2, containing 15662 OB stars with {\it Gaia}-DR2  data. Its distribution in the XY plane is shown in the mid panel of the left column of Fig. \ref{fig:mapesxy}. The other is the Xu et al. catalogue (\cite{xu}), containing 9750 OB stars and based on {\it Gaia}-eDR3 parallaxes and spectroscopic data. Its XY distribution is shown in the bottom panel of the left column in Fig. \ref{fig:mapesxy}. Neither of the catalogues has the radial velocities of the objects. For some of them, however, this information can be obtained through SIMBAD Database. 

\subsection{Analysis of the distribution in the galactic plane}
\label{xysection}

In Fig. \ref{fig:mapesxy} we compare and analyze the spatial distribution of all these tracers. Their location has been projected on the Galactic plane covering an area of $\pm 3 kpc$ from the position of the Sun. In the left column of this figure we report, from top to bottom, the distribution of the OCs, the OB field stars by \cite{pantaleoni} and the ones from Xu et al. \cite{xu}, respectively. At the right panels we have the same maps but with the CCs oveplotted.
\newline

A clear correlation between the CCs and the three stellar tracers’ population is clearly seen.
The inclined straight grey lines delineate the region containing the RW in the XY map as proposed by Alves et al. \cite{alves}. Both lines are parallel and tilted 30º relative to the Y axis in the first galactic quadrant, one passing through the Sun’s position (origin) and the other being horizontally translated 600pc to the left. We observe that in the first and second galactic quadrants the Radcliffe Wave structure is almost parallel to these lines indicating that the stellar and gas components are almost centred in the middle of the delineated region. On the contrary, in the third quadrant, it seems that the gaseous CCs  are more tilted towards the X axis, thus toward the anticenter direction, with an inclination in the range $40-50\deg$ respect to the Y axis.
\newline
\newline
\newline
Whereas this different inclination is also observed in the open cluster sample, we can not confirm it when using the OB field stars.  This slightly different inclination of the structure should be analyzed in detail (work in progress).

From Fig.\ref{fig:mapesxy} we also realize that our three samples of stellar tracers have objects in the delineated region of the Radcliffe Wave which are further than the distances reached by the CC catalogue. We wonder about a possible more spatially extended characterisation of the RW structure from the stellar component. In the present work, and for simplicity, we focuse only on tracers which have a physically linked CC.

\section{Methodology}

The study of the Radcliffe Wave structure consists of three parts. The first one has been the evaluation of its spatial distribution in the XY galactic plane, analysis that has been done in the previous Section. As a second step, we aim to map the vertical distribution along the RW axis, defined approximately as the line following the structure (the grey line passing through the Sun’s position in \ref{fig:mapesxy}) and, as a third step, and for the first time, we aim to analyze the kinematic behaviour of its members.  To address these steps, a positional cross-match needs to be done between our stellar tracers and the CCs defining the RW. This work is presented in Sect. \ref{link}. The study of the dynamical evolution of the structure requires to use the code galpy for the integration of the tracers' orbits back on time in a realistic Galactic potential. The methodology is presented in Sect. \ref{orbit}. In this work the methodology to carry out both studies has been developed, and it has been applied to obtain the  association of some OCs to the CCs. The same analysis but looking for the link between the CCs and the OB field stars using both catalogues as tracers is deferred to a future study.

\subsection{Link of the stellar tracers to the Cloud complexes}
\label{link}

\begin{figure}[h]
\centering
\includegraphics[width=8.5cm]{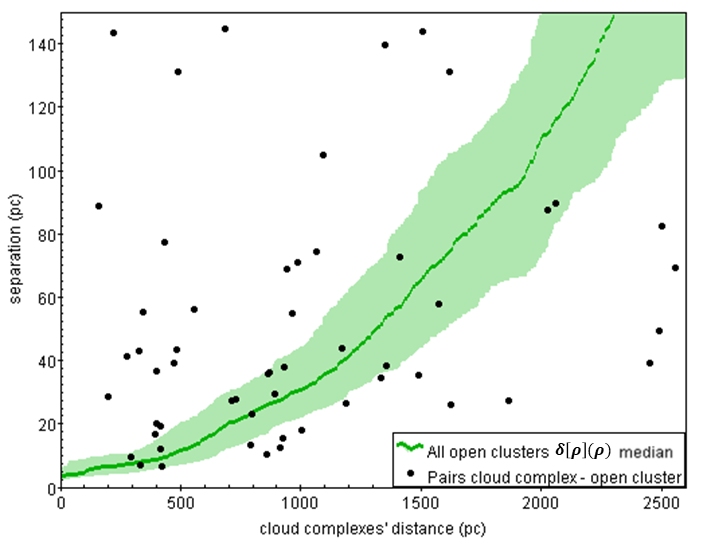}
\caption{Separation of an OC-CC pair as a function of the CC’s distance (for OCs younger than 30Myr).The green solid line is the median of the OC’s distance uncertainty derived in Appendix B (see Fig. \ref{figapendix2} bottom). The region shaded in green includes this error with a probability of 68\%, because it is delineated by the 84th percentile at its upper part and the 16th percentile at its lower part.}
\label{fig:OCCCpair}
\end{figure}

In order to identify which of the 349 OCs of Castro-Ginard et al. catalogue (\cite{castro}) that are younger than 30Myr belong to the Radcliffe Wave, first a cross-match has been made between this sample and the Alves  et  al. catalogue (\cite{alves}) of Cloud Complexes. Using TOPCAT, we have matched OCs and CCs that are at distances less than 150 pc. To do so, a sphere of radius 150pc is centred at each OC and it is considered that this OC matches with all CCs which fall within the sphere. 55 pairs composed of one OC and one CC nearer than 150pc have been found. No OC is found to have two CCs inside its sphere.
Then, the separation between the objects of each pair has been compared to the uncertainties of their respective distance to decide whether they can be considered coupled or not. As discussed in APPENDIX B, the distance uncertainty that needs to be taken into account is the one associated with the OCs’ distance, because the OCs' distance uncertainty is always equal to or greater than the CCs' distance uncertainty (see Fig. \ref{figapendix2})). The analysis has been done by plotting each pair’s separation against the corresponding CC distance; and also representing the 50th percentile of the distribution of OC's distance uncertainty ($\delta[\rho]$) as a function of OC’s distance ($\rho$) (green line in Fig. \ref{fig:OCCCpair}), as well as colouring the region delimited by its 16th and 84th percentiles (green shaded region in Fig. \ref{fig:OCCCpair}). In this Fig. \ref{fig:OCCCpair}, we can clearly compare for each OC-CC pair the magnitude of their separation with the  error in the OC’s distance estimation at a given distance.

A criterion must be established to decide which pairs are considered actually coupled and which ones are considered not enough close to be physically linked. Being conservative about the uncertainty distance might have, it has been chosen to consider the pair coupled when their separation is below the 84th percentile of the OC’s distance error, i.e., if the point in Fig. \ref{fig:OCCCpair} lies below the upper limit of the green shaded region.

\begin{figure}[t]
    \centering
    \includegraphics[width=8.5cm]{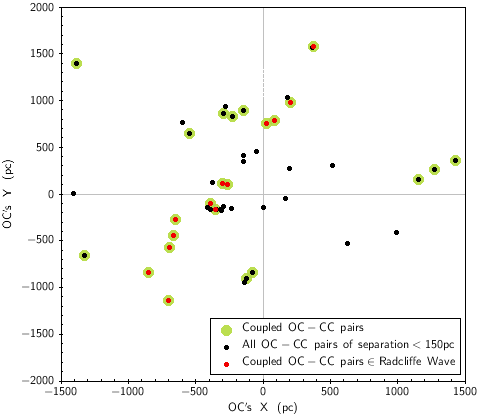}
    \caption{All points are positions of OCs younger than 30Myr which have a CC at less than 150pc. The ones centred in green circles correspond to physically coupled pairs of one OC and one CC. Among these, the 13 in red colour are paired OC-CC which belong to the region where the Radcliffe wave is defined. A few distant outliers have been left out of the figure.}
    \label{fig:OCCC-RW}
\end{figure}

Applying this strategy, 31 out of the total 55 OC-CC pairs less than 150 pc apart are found to be coupled. However, they do not all belong to the RW as is defined in Fig. 2 of Alves et al. paper (\cite{alves}). Following this paper figure, only the coupled OC-CC pairs which visually fall within the narrow XY region of approximately 2.7 kpc of length defined as the RW are selected (delineating this region with two straight parallel lines as described in Fig. \ref{fig:mapesxy}). By applying these limits,  a sample of 13 OC-CC pairs belonging to the RW is obtained. Their location is shown in Fig. \ref{fig:OCCC-RW}. As the kinematic data are available for the OCs, by means of their orbit integration the movement of their associated Radcliffe Wave’s cloud complexes can also be traced.

\subsection{Trace-back orbital analysis }
\label{orbit}

We want to characterize not only the spatial distribution of the Radcliffe Wave, but also its dynamical evolution. With that, we aim to evaluate  whether its constituents really describe a spatially and kinematically coherent structure.

In order to trace back the orbit of the RW tracers, the Python galpy package for galactic dynamics is used. Providing the initial ra, dec, distance, proper motion in ra and in dec, and the heliocentric line-of-sight velocity, it provides the Galactocentric-Cartesian coordinates $X$, $Y$, and $Z$  \footnotetext{Output Galactic coordinates of the Galpy code: $X$ axis is in the Sun - Galactic center direction, centered at the Galactic Center and positive toward the Galactic Anticenter; $Y$ axis positive toward the Galactic rotation and $Z$ perpendicular to the Galactic plane} as well as their respective velocities $V_{X}$, $V_{Y}$, and $V_{Z}$ at any time of the orbit integrated in a certain Galactic potential. For this first study, the chosen potential has been \emph{MWPotential2014} (\cite{bovy}), a simple model for the Milky Way’s gravitational potential. It is a 3D axisymmetric potential composed by a spherical bulge, a Miyamoto-Nagai potential for the disk and a NFW (Navarro-Frenk-White) potential for the dark matter halo.
\newline
The orbit integration for the RW tracers needs to go back in time, after the established initial conditions, only up to few tens of Myr, the maximum quantity assumed for the Radcliffe Wave’s age. We want to emphasize that in this so short a time a linear trace-back on time using the current velocity vector would be almost equivalent. The galpy package for orbit integration has been used here only to get familiarized with the code, although it is not strictly necessary for the study.

\section{Results}

In order to visualize the undulation of the RW structure traced by the 13 OC-CC pairs identified in Fig. \ref{fig:OCCC-RW}, an axis $y_{prime}$ is defined as follows: 
\begin{equation}
  y_{prime} = y \cos({30}^{\circ})+x \sin({30}^{\circ}) 
\end{equation}
Where $(x, y, z)$ are the heliocentric Galactic coordinates and the $y_{prime}$ axis corresponds to the straight line passing through the Sun’s position in Fig. \ref{fig:mapesxy}, and is approximately lined up with the RW structure in the first and second galactic quadrants, as done in Alves et al. paper (\cite{alves}). As mentioned in Sect. \ref{xysection}, the inclination of the RW could differ in  the  third  Galactic quadrant with respect to what is observed in the first and second quadrants. This misalignment is not considered here for two reasons. Firstly, we want to reproduce Alves et al. (\cite{alves}) paper’s steps as much as possible, and secondly, a change in the inclination just at the Sun location is suspicious, why just at Sun's location?

\begin{figure*}[htb!]
   \subfloat[]{\includegraphics[width=0.9\textwidth]{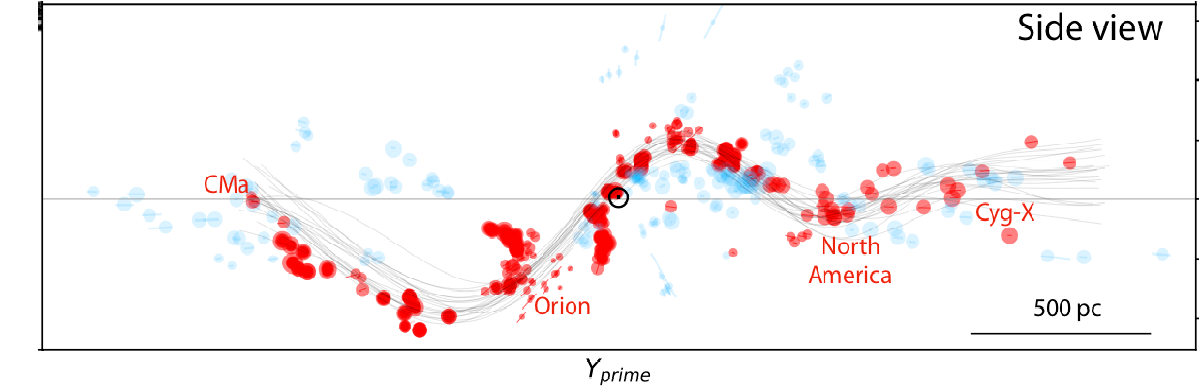}}
  \vskip\baselineskip
   \subfloat[]{\includegraphics[width=0.92\textwidth]{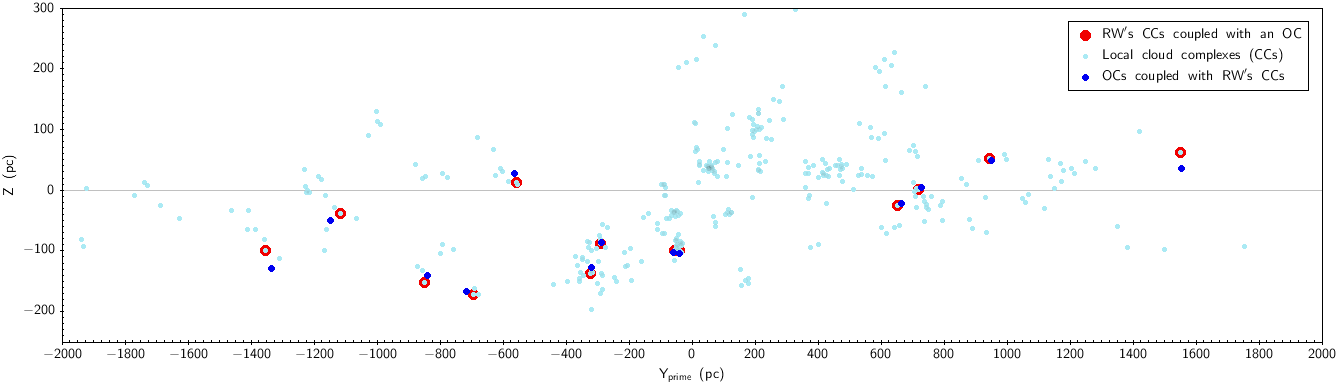}}
   \vskip\baselineskip
   \subfloat[]{\includegraphics[width=0.98\textwidth]{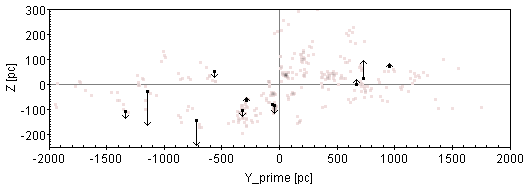}}
   \caption{(a): the RW as published by Alves et al. \cite{alves} (see their Fig. 2 for explanation). (b): the RW  as seen by cloud complexes and young open clusters. The light blue points are all the local cloud complexes in Alves et al. catalogue (\cite{alves}). Among these, the 13 with red circles belong to the RW and have an associated OC, whose position is represented as a dark blue point (these are the 13 OC-CC pairs identified in Fig. \ref{fig:OCCC-RW}). (c): The Galactic velocity perpendicular to the Galactic plane ($V_Z$) for the 11 OC-CC pairs is shown as black arrows (see text for explanation). }
    \label{figVz}
\end{figure*}

In Fig. \ref{figVz} (mid panel) we present the distribution of the 13 selected CC-OC pairs in the ($y_{prime}$, $z$) plane, i.e., approximately along the RW structure.  This figure bears a close resemblance to Fig. 2 in Alves et al. paper (\cite{alves}), reproduced in the top panel in Fig. \ref{figVz} for comparison. We observe that the 13 CCs which have been found to be linked to a young OC are quite uniformly distributed along the wave. Although this small number of objects has very little statistical significance, from now on we will assume that this small sample is an acceptable representation of the CCs defining the RW.

\begin{figure*}[htb!]
   \subfloat[]{\includegraphics[width=0.5\textwidth]{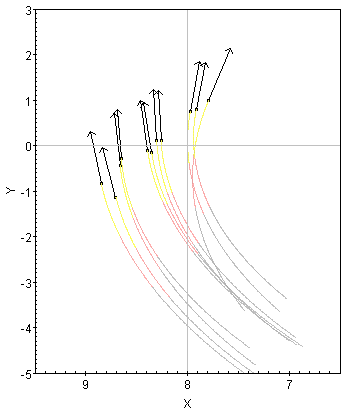}}
   \subfloat[]{\includegraphics[width=0.5\textwidth]{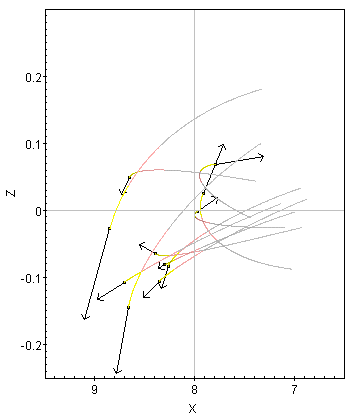}}
   \caption{Orbital motion of the 11 OC-CC pairs belonging to the Radcliffe Wave identified in Fig. \ref{fig:OCCC-RW} and highlighted in Fig. \ref{figVz} (b). The black arrows represent the current velocity vector of the OC of each pair. Each curve describes the orbit of an OC-CC pair, highlighted in different colours as a function of time: the yellow part corresponds to the trajectory from the present to 5Myr ago, the orange part corresponds to the trajectory comprised between 5Myr and 10Myr ago, and the gray part corresponds to the trajectory comprised between 10Myr and 20Myr ago. (a): projection on the Galactic plane, with the galactic center located at $(X,Y) = (0,0)$ kpc and the Sun at $(X,Y) = (8.0,0)$ kpc (b): projection on the $(X,Z)$ meridional plane. }
    \label{figXYZ}
\end{figure*}

Out of this sample of 13 OC-CC pairs belonging to the Radcliffe Wave, the radial velocity is known for 11 OCs. In Fig. \ref{figVz} (bottom panel) we plotted with black arrows the velocity component perpendicular to the Galactic plane ($V_Z$) of these CC-OC pairs. From a simple model of harmonic motion in the vertical direction we should expect a maximum velocity of the particles when crossing the plane and an almost null vertical velocity at the positions of maximum elongation, that is, when the separation from the  $ z = 0$ plane is maximum. Observing the obtained distribution of velocities it is difficult to corroborate or refuse this simple wave model. We observe how some of the CC-OC pairs, in particular the Cloud Complex of  Monoceros R2, at the position $(y_ {prime}, z)$ = (-720, -170) pc and associated with the open cluster NGC 2183, is located at the maximum separation from the plane  ($z=0$) according to the models presented by Alves et al. \cite{alves} and presents one of the largest velocity perpendicular to the plane. Undoubtedly, more kinematic information is needed to confirm or refuse the existence of this wave.

The orbit integration back in time for the 11 CC-OC pairs is presented in Fig. \ref{figXYZ}. We show both, the projected orbit in the galactic plane (left panel) and on the vertical meridional plane (right panel). As described in Sect. \ref{orbit},  the \emph{MWPotential2014} potential has been used in this trace-back orbital analysis. As expected, the orbits of some of the CC-OC pairs seem to have a common origin. This is the case of the OC-CC pairs associated to the Orion star forming region, located around $(X,Y) \sim (8.3, -0.07) $ kpc. They present a clear concentration of their orbits (minimum dispersion in the $X,Z$ plane) about 5-10 Myr ago. However, as a general trend, our results indicate that the orbits of the CC-OCs pairs do not suggest a common origin for the full set of members associated with the wave. If a common origin would be proposed for this RW structure, its origin would neither be associated to a point nor to a straight line in the $XZ$ plane. Further work is needed to project this orbits in a plane parallel or perpendicular to the $y_{prime}, z$ plane (work in progress).

\section{Conclusions and work in progress}

In this work we present a first attempt to characterize the stellar component of the Radcliffe Wave. We have considered as tracers the most updated catalogues of young open clusters and the OB field stars, both with accurate {\it Gaia} DR2 and eDR3 data. As expected, the location of the young open clusters with ages up to 30 Myr shows a high correlation with the location of the Cloud Complexes used to discover and characterize the Radcliffe Wave.  Being conservative, we have identified 13 open clusters physically linked to a Cloud Complex as probable members of the Radcliffe Wave. This validates the use of them as tracers of this structure.

As seen in Alves et al. \cite{alves} the projection on the Galactic plane of the Cloud Complexes belonging to the RW seems to present an elongated structure that departs from a straight line when crossing from the first and second Galactic quadrant to the third one. This behaviour is suspicious since the Sun should not be expected to have a privileged location in relation to this structure. Although more work should be devoted in the future, our selected OB field stars do not seem to confirm it. A Kernel non parametric analysis is in progress to statistically define the location of these under or over densities in the plane. Our $XY$ maps for the OB field stars indicate that accurate distance information is available for stars that are further than the distances reached by the CC catalogue. This encourages us to use this sample to confirm or refuse the existence of the Radcliffe Wave.

The link performed here between the Cloud Complexes and the Open clusters has allowed us, for the first time, to look at the kinematic behaviour of 11 CC-OCs members associated with the Radcliffe Wave. From the analysis of both the vertical motion and the integration backwards in time, though it is not conclusive, we can sum-up that: 1)  the velocities are not contradictory with the behaviour expected from a simple model of harmonic motion in the vertical direction, and 2)  the orbits back on time neither suggest an origin  associated to a point nor to a straight line in the $XZ$ plane. Work is in progress to project the orbits in the planes  parallel and perpendicular to location proposed for the Radcliffe Wave.

A strategy similar to the one developed here to link the Cloud Complexes to the Open Clusters is being designed to link the OB field stars population to the at present components associated with the Radcliffe Wave. The kinematic information associated with this link will be of crucial importance to corroborate or refute the existence of this structure.  We plan to repeat our work by using a larger sample of tracers, but the scheme developed seems a useful tool for providing new insights on the RW. Our first analysis indicates that there are sufficient proper quality data, and it is expected that a larger sample of CC and tracer pairs belonging to the RW will be obtained with respect to the 13 OC-CC pairs identified in this study.




\vspace{0.5cm}
\begin{acknowledgments}
We thank Dr. Jesús Maíz and Michelangelo Pantaleoni for kindly providing the Alma catalogue of OB stars (II); as well as Dr. Alfred Castro and Tristan Cantat-Gaudin for kindly providing the catalogue of open clusters.
We also thank Dr. Núria Miret for her collaboration, and Juan Carbajo. This work was (partially) funded by the Spanish MICIN/AEI/10.13039/501100011033 and by "ERDF A way of making Europe" by the “European Union” through grant RTI2018-095076-B-C21, and the Institute of Cosmos Sciences University of Barcelona (ICCUB, Unidad de Excelencia ’Mar\'{\i}a de Maeztu’) through grant CEX2019-000918-M.
\end{acknowledgments}


\clearpage

\begin{center}
\textbf{APPENDIX}
\end{center}

\begin{center}
\textbf{APPENDIX A: COMPARISON OF DISTANCE ESTIMATES}
\end{center}

In this section we draw a comparison between distance estimates for OB stars of the Pantaleoni-González et al. catalogue (\cite{pantaleoni}), which are used as tracers.

\begin{figure}[hbt!]
   \subfloat[ALS2 vs {\it Gaia}]{\includegraphics[height=3.9cm]{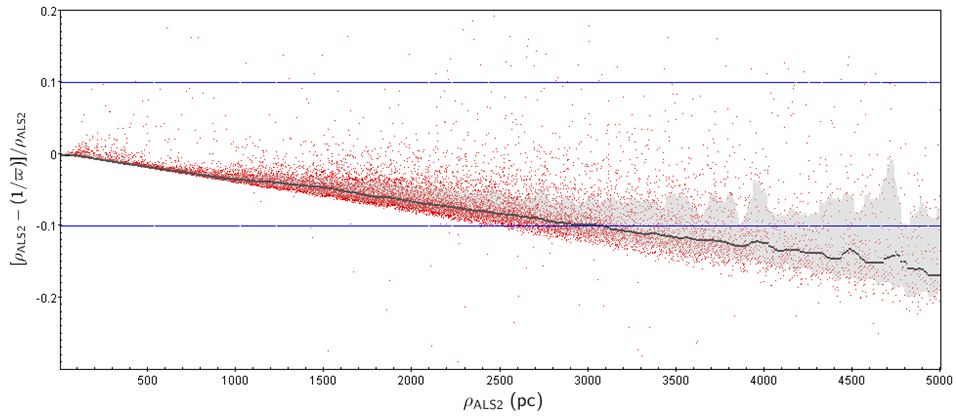}}
  \vskip\baselineskip
   \subfloat[ALS2 vs StarHorse2]{\includegraphics[height=3.9cm]{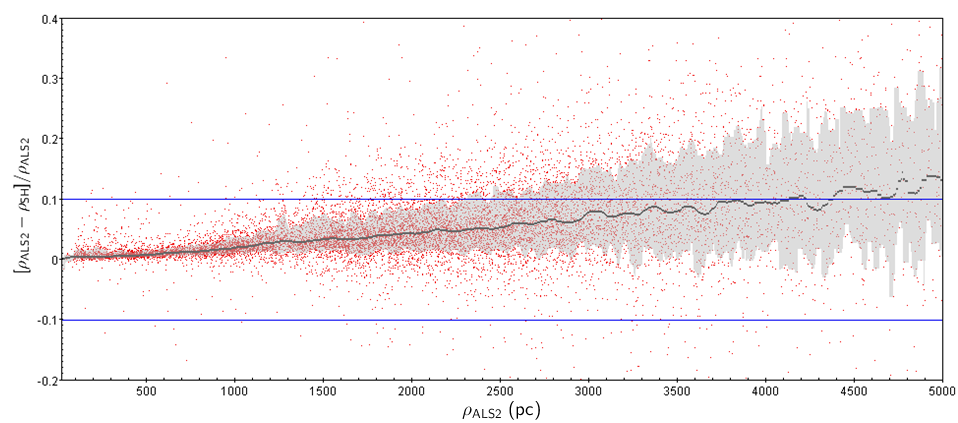}}
   \caption{Relative differences on stellar distances when comparing the distance estimation using J. Maíz Apellániz's prior (\cite{maiz2001} and \cite{maiz2005}) with (a) the estimations inverting  the observed parallax, and (b) using F. Anders’s prior (\cite{anders2019}).}
    \label{figapendix1}
\end{figure}

Their distance can be estimated inverting the observed parallax, or else more precisely through two Bayesian formalisms mentioned in (\cite{pantaleoni}): using J. Maíz Apellániz's prior (\cite{maiz2001} and \cite{maiz2005}) or using F. Anders’s prior (\cite{anders2019}). We study the relative difference between the estimation using J. Maíz Apellániz's prior and inverting the observed parallax, as well as the relative difference between the estimations using both priors. As seen in Fig. \ref{figapendix1}, inverting the observed parallax results in distances systematically greater than the ones estimated using J. Maíz Apellániz's prior; while distances estimated using F. Anders’s prior are systematically smaller than the ones estimated using J. Maíz Apellániz's prior. However, in both cases the relative differences reach 10\% at distances farther than 3kpc. Therefore, for the purposes of this study the amount they differ at near distances is not significant, so the distance estimated using J. Maíz Apellániz's prior is taken as the distance of the sample of OB stars of the Pantaleoni-González et al. catalogue (\cite{pantaleoni}).
\newline

\begin{center}

\textbf{APPENDIX B: DISTANCE UNCERTAINTIES}
\end{center}

In this section we present the distance uncertainties of Cloud Complexes and Open Clusters.

\begin{figure}[h]
   \subfloat[Cloud Complexes]{\includegraphics[height=6.7cm]{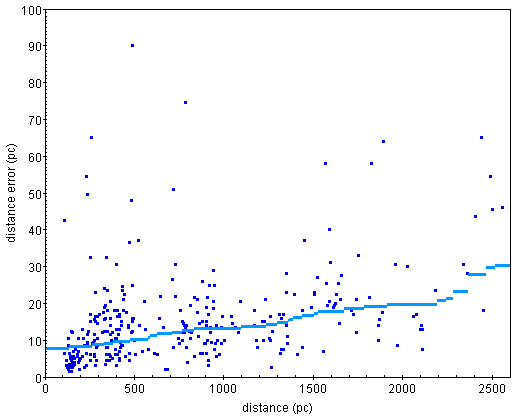}}
  \vskip\baselineskip
   \subfloat[Open Clusters]{\includegraphics[height=6.7cm]{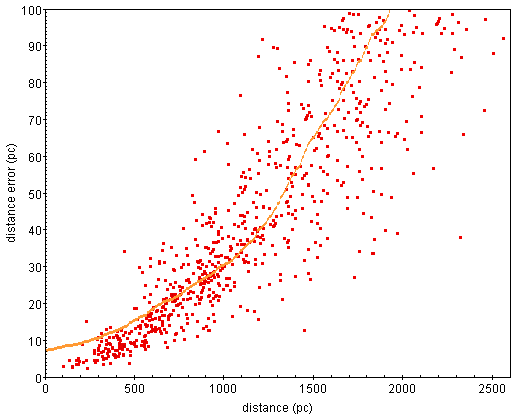}}
   \caption{Distance uncertainty for the Cloud Complexes (top) and Open Clusters (bottom) as a function of distance. We have left a few distant outliers out of the plot. The solid line is the 50th percentile partially smoothed.}
    \label{figapendix2}
\end{figure}

The error of the CCs' accurate distance estimations is considered to be a symmetric error bar which includes the central value (the median: 50th percentile) with a probability of 68\%. So, this error bar is defined by the extreme values corresponding to the 16th and 84th percentiles; although there is no reason for this error to be symmetric. Its behaviour as a function of the CC's distance is shown in the top panel (a) of Fig. \ref{figapendix2}.

As a first approach, the distance of the OCs in Castro-Ginard et al. catalogue (\cite{castro}) is calculated using their parallax $\varpi \pm \Delta{\varpi}$, as $\rho \pm \Delta (\rho) = \frac{1}{\varpi} \pm \frac{\Delta (\varpi)}{\varpi^2}$. Although this estimator is biased, for the near distances studied this effect is not the most limiting one and it is relatively safe and justified to use it. The behaviour of this estimated distance uncertainty as a function of OC's distance is shown in the bottom panel (b) of Fig. \ref{figapendix2}.

The error in both distances follows a tendency to increase with distance (as seen in Fig. \ref{figapendix2}); but, having both vertical axes the same scale, it is clear that the dominant uncertainties are the ones associated with the OCs (as their 50th percentiles are equal below 200pc, but for farther distances OCs’ 50th percentiles is above the CCs’ one).



\clearpage

\end{document}